\documentclass[twoside]{mhd}
\usepackage{amsmath}
\usepackage{graphicx}
\usepackage{caption}
\usepackage{subcaption}
\usepackage{multirow}
\usepackage{float}
\usepackage{bm}
\usepackage{textcomp}
\mhdhead{40}{1}{1}

\newcommand\T{\rule{0pt}{2.6ex}}       % Top strut
\title{Numerical and Experimental Investigation of Electro-Vortex Flow in a Cylindrical Container}

\author{K.~Liu\inst{1,2*}, F.~Stefani\inst{1}, N.~Weber\inst{1}, T.~Weier\inst{1}, B.W.~Li\inst{2}}

\institute{Helmholtz-Zentrum Dresden-Rossendorf, Bautzner Landstra{\ss}e 400, 01328 Dresden, Germany 
	\and School of Energy and Power Engineering, Dalian University of Technology, Dalian, 116024, China
          }

\begin{document}
	\maketitle
	
% ------------------------------ Abstract --------------------------------
	\begin{abstract}
		In a cylindrical container filled with an eutectic GaInSn alloy, an electro-vortex flow (EVF) is generated by the interaction of a non-uniform current with its own magnetic field. In this paper, we investigate the EVF phenomenon numerically and experimentally. Ultrasound Doppler Velocimetry (UDV) is applied to measure the velocity field in a cylindrical vessel. Second, we enhance an old numerical solver by taking into account the effect of Joule heating, and employ it for the numerical simulation of the EVF experiment. Special focus is laid on the role of the magnetic field, which is the combination of the current induced magnetic field and the external geomagnetic field. For getting a higher computational efficiency, the so-called parent-child mesh technique is applied in OpenFOAM when computing the electric potential, the current density and the temperature in the coupled solid-liquid conductor system. The results of the experiment are in good agreement with those of the simulation. This study may help to identify the factors that are essential for the EVF phenomenon, and for quantifying its role in liquid metal batteries.
		
	\end{abstract}

% ---------------------------- Introduction ------------------------------
	\section{Introduction}
The object of the present study is the vortex flow that appears due to the 
electro-magnetic force in a cylindrical container, which is filled with 
the liquid metal alloy GaInSn. The considered flow is commonly known as electro-vortex 
flow (EVF), which develops at (or close to) a changing cross-section of a 
liquid conductor \cite{Shercliff1970}. The changing cross-section deforms the electric currents 
and the distribution of the induced magnetic field and modifies, 
thereby, the intensity and 
direction of the Lorentz force. In our physical model, see Fig. \ref{fig:experimental_set-up}\hspace{1.5pt}\;b, there is a sharp 
change of cross-section between the top copper electrode and the fluid zone 
(light blue colour). The arising 
Lorentz force is non-conservative, i.e., its curl is not equal to zero. Since 
such a force
cannot be balanced completely by a pressure gradient, 
it immediately drives a fluid flow
(since the EVF is not an instability, it does not need any critical threshold
to set in). By comparing Fig. \ref{fig:experimental_set-up}\hspace{1.5pt}\;a and Fig. \ref{fig:experimental_set-up}\hspace{1.5pt}\;b, we see 
that this model can also be considered as the top part of a liquid metal battery (LMB) \cite{Weber2014b}, 
in which EVFs should be considered carefully, both
as a possible danger for the integrity of the liquid electrolyte \cite{Stefani2015,Herreman2019b}, as well as for
their potentially positive role in enhancing mass transfer in the electrodes \cite{Kelley2018,Ashour2017a,Weier2017,Weber2018}. Besides 
its relevance for LMBs, 
EVF is a well-known phenomenon in quite a number of engineering devices such as  
DC and AC electric arc furnaces \cite{Kazak2013}, induction 
furnaces \cite{Kazak2012}, electric jet engines 
\cite{Kazak2011} or pumps \cite{Denisov2016} -- for an extensive overview, see \cite{Bojarevics1989}.

In this paper, we study  a paradigmatic model of an EVF, both experimentally and 
numerically. In our  experiment,  
we employ (because of the opaqueness of GaInSn) Ultrasound Doppler Velocimetry (UDV) to measure the liquid metal's 
velocity. As shown in Fig. \ref{fig:experimental_set-up}\hspace{1.5pt}\;c, four 
UDV sensors (1, 3, 5, 7) are used at the same time for measuring the transient 
velocity of the fluid. With this non-invasive technique, we obtain detailed flow field information about the EVF, which can be processed and compared with corresponding
numerical results. To conduct the numerical simulation of this EVF experiment, the 
solver by Weber et al. \cite{Weber2017b} was extended by considering the Joule heating effect. Just as in the previous solver, the electric 
potential, the current density and the temperature are solved on 
a parent mesh, while the other variables, such as velocity and pressure, 
are solved only in the fluid zone \cite{Beale2016}. The parent mesh, however,
covers all conductive parts as shown in Fig. 
\ref{fig:experimental_set-up}\hspace{1.5pt}\;b.

The paper is structured as follows:  A brief description of the experimental set-up is 
given in the following section. The numerical model with the employed 
boundary and initial conditions is discussed in section 3. Section 4 presents the numerical results 
and their comparison with the measured data. Some conclusions 
can be found in Section 5.
    
% ----------------------------- Section 2 --------------------------------
    \section{The experimental set-up}

\begin{figure}[htbp]
     	\centering
     	\begin{subfigure}[b]{0.175\textwidth}
     		\centering
     		\includegraphics[width=\textwidth]{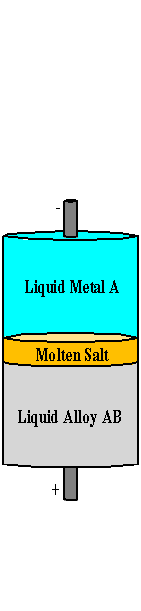}
     		\captionsetup{skip=10pt}
     		\caption{}
     		%\label{fig: parabola surface fit origin 4p contour}
     	\end{subfigure}
     	\hfill
     	\begin{subfigure}[b]{0.445\textwidth}
     		\centering
     		\includegraphics[width=\textwidth]{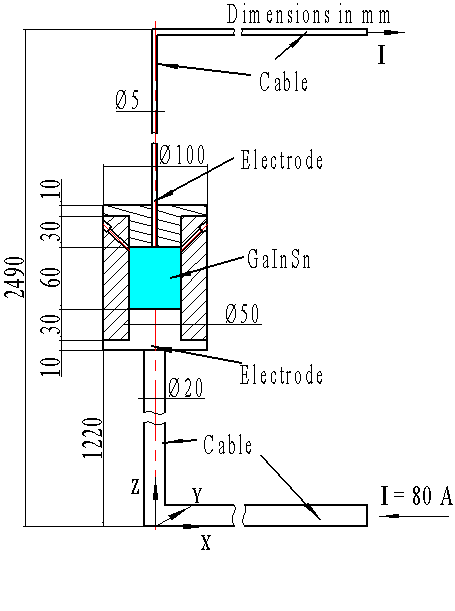}
     		\captionsetup{skip=10pt}
     		\caption{}
     		%\label{fig: parabola surface fit origin 4p curve}
     	\end{subfigure}
        \hfill
        \begin{subfigure}[b]{0.345\textwidth}
        	\centering
        	\includegraphics[width=\textwidth]{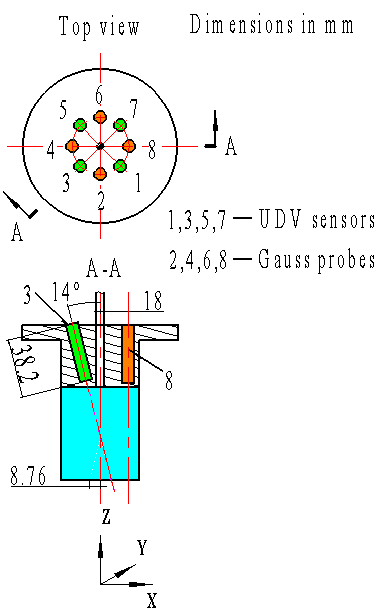}
        	\captionsetup{skip=10pt}
        	\caption{}
        	%\label{fig: parabola surface fit origin 4p curve}
        \end{subfigure}
     	\caption{A schematic of an LMB (a), and the EVF experiment: (b) Sketch of the experiment. 
	(c) Partially enlarged (1.5:1) cross-sectional view showing the distribution of the UDV sensors and the holes for the magnetic field measurements in more detail.}
     	\label{fig:experimental_set-up}
     \end{figure}

A schematic of the experimental set-up is shown in Fig. \ref{fig:experimental_set-up}\hspace{1.5pt}\;b. This set-up consists of six main parts: one top cable, a top electrode, a cylindrical container filled with 
liquid GaInSn, a bottom electrode, a bottom cable, and a current source 
(not shown in Fig. \ref{fig:experimental_set-up}). Note that both the 
top and the bottom cable are bent at a rather large distance from the 
vessel (appr. $1.2$\;m).
The lid and the side wall of the cylindrical container, 
whose inner radius is $R=25$\;mm, are made of plastics. 
In order to completely fill the whole container and to avoid any air pockets, 
there are two inclined small holes in the cylindrical side wall, close to its connection with the top bottom of the lid. The top electrode, whose diameter 
is $5$\;mm, has the same centre line as the lid of the container. However, 
the bottom electrode has the same diameter ($50$\;mm) as the inner 
diameter of the container, so that no EVF is expected to set in there. A setup similar to our experiment was already studied experimentally using fiber-optic velocity measurement by Zhilin \cite{Zhilin1986,Volokhonskii1991}. Moreover, several theoretical \cite{Butsenieks1976,Chudnovskii1989b,Millere1980,Vlasyuk1987b,Bojarevics1989,Kolesnichenko2005} and numerical \cite{Nikrityuk2007,Weber2017b,Herreman2019b} works of this generic geometry are available in the literature.

A DOP 3010 Ultrasound Doppler Velocimetry (UDV) from Signal-Processing was used to measure liquid metal's velocity.  
Figure \ref{fig:experimental_set-up}\hspace{1.5pt}\;c indicates the positions and inclinations
of the holes where the UDV sensors (1, 3, 5, 7) are installed. 
Additionally, we measure the Earth's magnetic field within the container 
(Table \ref{tab:the_measured_magnetic_field}) by means of
a Lakeshore Gauss meter. 
Figure \ref{fig:experimental_set-up}\hspace{1.5pt}\;c shows the 
positions 2, 4, 6, 8 for the measurement of the
external magnetic fields (specify model just for consistency).
   
    \begin{table}[htbp]
    	\centering
    	\caption{The measured geo-magnetic field}
	    \begin{tabular}{p{9.565em}ccc}
	    	\hline \T
	    	Position/Mean value & \multicolumn{1}{p{4.19em}}{$b_x \hspace{1.5pt}$ (mT)} & \multicolumn{1}{p{4.19em}}{$b_y \hspace{1.5pt}$ (mT)} & \multicolumn{1}{p{4.19em}}{$b_z \hspace{1.5pt}$ (mT)} \\
	    	\hline \T %\vspace{0.6mm}
	     	$P_2$    & 0.0144 & 0.0333 & -0.0133 \\
	    	$P_4$    & 0.0164 & 0.0336 & -0.0161 \\
	    	$P_6$    & 0.0152 & 0.0338 & -0.0249 \\
	    	$P_8$    & 0.0146 & 0.035 & -0.0097 \\
	    	\hline \T
	    	Mean value & 0.0151 & 0.0339 & -0.0160 \\
	    	\hline
	    \end{tabular}%
        \label{tab:the_measured_magnetic_field}%
    \end{table}

    As shown in Table \ref{tab:the_measured_magnetic_field}, the $b_x$ and $b_y$ values of the Earth's magnetic fields at the different positions are almost the same. So, it seems well justified to use their average value as 
   the external field in the entire volume of the vessel. The issue is a bit more complicated
   for $b_z$ which shows some larger deviations from its average value 
   of $-16$\;$\mu$T. Those might be attributed to some remaining 
   permeability of the (otherwise austenitic) screws in the lid.
   An additional problem comes up
   when the current source is switched on \cite{Starace2014}. 
   While the arising azimuthal component of the mainly axial current 
   can reliably be computed by Biot-Savart's law (see next section), things are 
   different for $b_z$. Various test measurements 
   have shown that this component is not exactly zero, as it would be
   expected from a perfectly axial current distribution.
   At a current of $80$\;A, 
   the additional $b_z$ component 
   is in the order of a few $\mu$T, which seems to result from 
   field deforming effects of the contact of the upper cable to 
   the electrodes, perhaps in combination with the mentioned 
   permeability of the screws.
   Unfortunately, a precise $b_z$ measurement 
   turned out to be extremely challenging, not least because of the
   strong geometric sensitivity of $b_z$ measurements in the presence 
   of a much larger azimuthal field. Given the importance of the
   $b_z$ component in the area close to 
   the upper lid for the formation of the EVF, we will 
   resort to  some indirect inference method, by testing numerically 
   a variety of deviations from the observed 
   average value of $-16$\;$\mu$T.

% ----------------------------- Section 3 --------------------------------
    \section{Mathematical model and boundary conditions}

    In the mathematical model of our experimental set-up, the following assumptions are made: \\
    1. The magnetic permeability of all parts (including air) is assumed to be the value of vacuum.\\
    2. 	The two plastic parts (cylindrical rim and upper lid) are considered to be adiabatic for heat transfer.\\ 
    3.	The physical properties of liquid metal (viscosity, electric conductivity and thermal conductivity, etc.) are assumed to be homogeneous, isotropic and not to depend on the temperature.\\ 
    The flow in an incompressible, viscous and electrically conducting fluid is described by the Navier-Stokes equation (NSE) \cite{Weber2013}
    \begin{equation}
     \dot{\textbf{\textit{u}}}+(\textbf{\textit{u}}\cdot\nabla)\textbf{\textit{u}}=-\frac{\nabla p}{\rho}+ \nu\Delta \textbf{\textit{u}}+ \frac{\textbf{\textit{f}}_L}{\rho}+\textbf{\textit{f}}_T
    \end{equation}
    and the continuity equation
    \begin{equation}
    \nabla \cdot \textbf{\textit{u}}=0,
    \end{equation}
    with \textbf{\textit{u}} denoting the velocity, \textit{p} the pressure, $\rho$ the density, $\nu$ the kinematic viscosity, $\textbf{\textit{f}}_L$ the Lorentz force and $\textbf{\textit{f}}_T$ the buoyancy.

    The trigger of the EVF is the Lorentz force\\
    \begin{equation}
    \textbf{\textit{f}}_L=\textbf{\textit{J}} \times \textbf{\textit{B}},
    \end{equation}
    with \textbf{\textit{J}} meaning the total current density and \textbf{\textit{B}} the total magnetic field. Note that the total magnetic field consists of two parts: the first part is the static magnetic field $\textbf{\textit{B}}_0$, generated by the applied current \textbf{\textit{I}} (or the corresponding current density $\textbf{\textit{J}}_0$); the second part is the external Earth's magnetic field $\textbf{\textit{b}}$. Although the second part ($\textbf{\textit{b}}$) is rather small, it must not be neglected because this part is multiplied with the large current density $\textbf{\textit{J}}_0$ and will cause swirling flow \cite{Davidson1999}. The effect of $\textbf{\textit{b}}$ 
will be discussed in the next section.

    Another driving force is buoyancy \cite{Meng2016,Rusche2002},\\
    \begin{equation}
    \textbf{\textit{f}}_T = gh \nabla(1-\beta(T-T_0))
    \end{equation}
    with $g$ denoting the standard gravity, $h$ the vertical distance from the top surface of GaInSn zone to the position where the buoyancy is computed, $\beta$ the thermal expansion coefficient, \textit{T} the temperature and $\textit{T}_0$ the reference temperature. Commonly, if the higher temperature domain is located in the upper region, buoyancy will have no effect to fluid flow. But in our set-up, the 
    high temperature domain is located close to the upper electrode where the current density, and hence
    the Joule heating is the highest. Therefore, the effect of buoyancy should be carefully considered \cite{Davidson2000a}.
    For computing the transient temperature field, an energy equation needs to be solved \cite{Luo2016,Personnettaz2018a} as
    \begin{equation}
    \dot{\textit{T}}+(\textit{\textbf{u}} \cdot \nabla)\textit{T}= \nabla \cdot (\frac{\lambda}{\rho \textit{C}_\textit{p}} \nabla \textit{T}) +\frac{1}{\rho \textit{C}_\textit{p}}\frac{\textbf{\textit{J}}^2}{\sigma},
    \end{equation}
    with $\lambda$ meaning the thermal conductivity, $\textit{C}_\textit{p}$ the constant pressure specific heat, $\textit{\textbf{J}}$ the current density and $\sigma$ the electrical conductivity.

    The main magnetic field ($\textbf{\textit{B}}_0$) can be obtained by using Biot-Savart's law \cite{Schmidt1999,Meir2004}
    \begin{equation}
    \textit{\textbf{B}}_0(r)=\frac{\mu_0}{4\pi} \int dV^{'} \frac{\textbf{\textit{J}}(\textbf{\textit{r}}') \times (\textbf{\textit{r}}-\textbf{\textit{r}}')}{\left| \textit{\textbf{\textit{r}}}-\textit{\textbf{\textit{r}}}' \right|^3} ,
    \end{equation}
    with $\mu_0$ denoting the vacuum permeability. Here, $\textbf{\textit{r}}$ means the local vector where $\textbf{\textit{B}}_0$ is calculated, and $\textbf{\textit{r}}'$ is the integration coordinate, which runs through the whole domain where $\textbf{\textit{J}}$ exist. For obtaining $\textbf{\textit{J}}$, we have to solve the Poisson-type equation for the electric potential as
    \begin{equation}
    \nabla \cdot (\sigma \nabla \varphi)=\nabla \cdot (\sigma \textbf{\textit{u}}\times \textbf{\textit{B}}).
    \end{equation}

    This entire problem is then solved with the following boundary and initial conditions:\\
    1. For the electric potential, the boundary condition at the end of the top and 
    the bottom cable is
    \begin{equation}
    \partial \varphi / \partial \textit{\textbf{n}}=-\textit{J}/\sigma,
    \end{equation}
    and at the insulating parts it is
    \begin{equation}
    \partial \varphi / \partial \textit{\textbf{n}}=0.
    \end{equation}
    2. For the temperature field on the boundary surrounded by a plastic container, we assume:\\
    \begin{equation}
     \partial T / \partial \textit{\textbf{n}}=0,
    \end{equation}
    on the boundary open to air (part of the top and bottom electrode) \cite{Vilums2011}:
    \begin{equation}
    T_b=fT_0+(1-f)T_c,\quad f=(1+\frac{\lambda_s}{\alpha_s \delta})^{-1},
    \end{equation}
    and for obtaining the absolute temperature rise conveniently, we can simply set the initial and the reference internal temperature field to $0$ in our simulations.\\ 
    3. For the hydrodynamic part, no-slip boundary conditions are used at all boundaries with solid walls. This means that all velocity components are set to zero:\\
    \begin{equation}
    u=0, v=0, w=0.
    \end{equation}
    Additionally, because the liquid metal is at rest before we switch on the current source, 
    the internal field ($\textbf{\textit{u}}$) can also be initialised simply as $0$. In the equations above, 
    $\varphi$ is the electric potential, $\textbf{\textit{n}}$ is the normal vector at the respective wall, $
    T_b$ is the temperature on the boundary, $T_c$ is the temperature at each cell centre next to the boundary, 
    $\lambda_s$ is the heat transfer coefficient of the copper electrodes, $\alpha_s$ is the heat exchange coefficient, 
    and $\delta$ is the distance from the cell centre to the cell face centre.

    For solving and updating the flow field, the PIMPLE algorithm, which combines the PISO algorithm and the SIMPLE algorithm together, is adopted to handle the pressure-velocity coupling. The Euler scheme is used for the temporal discretisation. The time step is adjusted automatically by comparing the Courant number and the magnetic diffusion Courant number with a preset value, respectively, when the current time step is finished. Therefore, it is not important how large the initial time step is set. At each time step, the solution is considered to be converged when the residual errors of the energy equation are less than $10^{-10}$, and the residual error of the momentum equation and the Poisson equation for the electric potential are less than $10^{-7}$, respectively.

% ----------------------------- Section 4 --------------------------------
	\section{Validation, comparison and discussion}

\begin{figure}[htbp]
		\centering
		\includegraphics[width=0.5\textwidth]{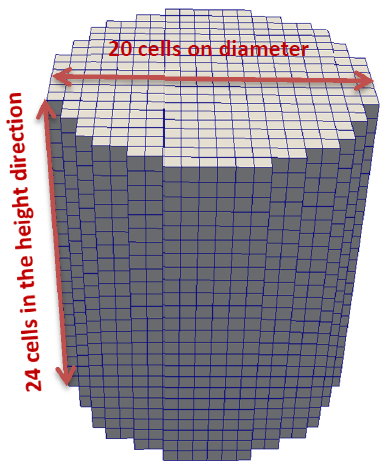}
		\caption{An example of the discretised of the fluid domain, here with the coarsest grid $20 \times 24$.}
		\label{fig:fluid_mesh}
	\end{figure}

	In order to set a complete case for our simulations, a 3D geometry was built in Inventor 2016, and saved as a STL file first. Then, we used \textit{snappyHexMesh} to discretise the 3D geometry up to millions of orthogonal uniform finite volumes. So, all smooth cylindrical surfaces will become small planes. A cross-section of the fluid domain is shown in \mbox{Fig. \ref{fig:fluid_mesh}}. The relevant physical parameters used in the simulations 
	are listed in \mbox{Tab. \ref{tab:physical_parameters}}. Most of the simulations (and experiments) 
	were done with an applied current of $80$\;A.
	
     % Table generated by Excel2LaTeX from sheet 'Sheet6'
     \begin{table}[htbp]
     	\centering
     	\caption{Physical parameters of liquid metal, cables and electrodes}
     	\begin{tabular}{lrl|l}
     		\hline \T
     		Parameters & Value & Dimension & Region \\
     		\hline \T
     		$\rho$   & 6403 & kg\;m$^{-3}$ & \\
     		$g$     & 9.8   & m\;s$^{-2}$  &  \\
     		$\lambda_s$ & 24 & W\,(m\,K)$^{-1}$ &  \\
     		$\sigma$ & 3.3$\times 10^6$ & S\,m$^{-1}$ & Fluid \\
     		$\mu_0$ & 1.25$\times 10^{-7}$ & N\,A$^{-2}$ & (GaInSn) \\
     		$\nu$    & 3.4$\times 10^{-7}$ & m$^2$\;s$^{-1}$ &  \\
     		$\beta$  & 1.2$\times 10^{-4}$ &   K$^{-1}$    &  \\
     		$C_p$  & 366 & J\,(kg\,K)$^{-1}$ &  \\
     		\hline \T
     		$\rho$   & 8960  & kg\;m$^{-3}$ & Cables \\
     		$\lambda_s$ & 401   & W\,(m\,K)$^{-1}$ &\hspace{3pt} and \\
     		$\sigma$ & 5.8$\times 10^{7}$ & S\,m$^{-1}$ & Electrodes \\
     		$C_p$  & 386   & J\,(kg\,K)$^{-1}$ &  \\
     		\hline
     	\end{tabular}%
     	\label{tab:physical_parameters}%
     \end{table}%

     \begin{figure}[htbp]
     	\centering
     	\begin{subfigure}[b]{0.485\textwidth}
     		\centering
     		\includegraphics[width=\textwidth]{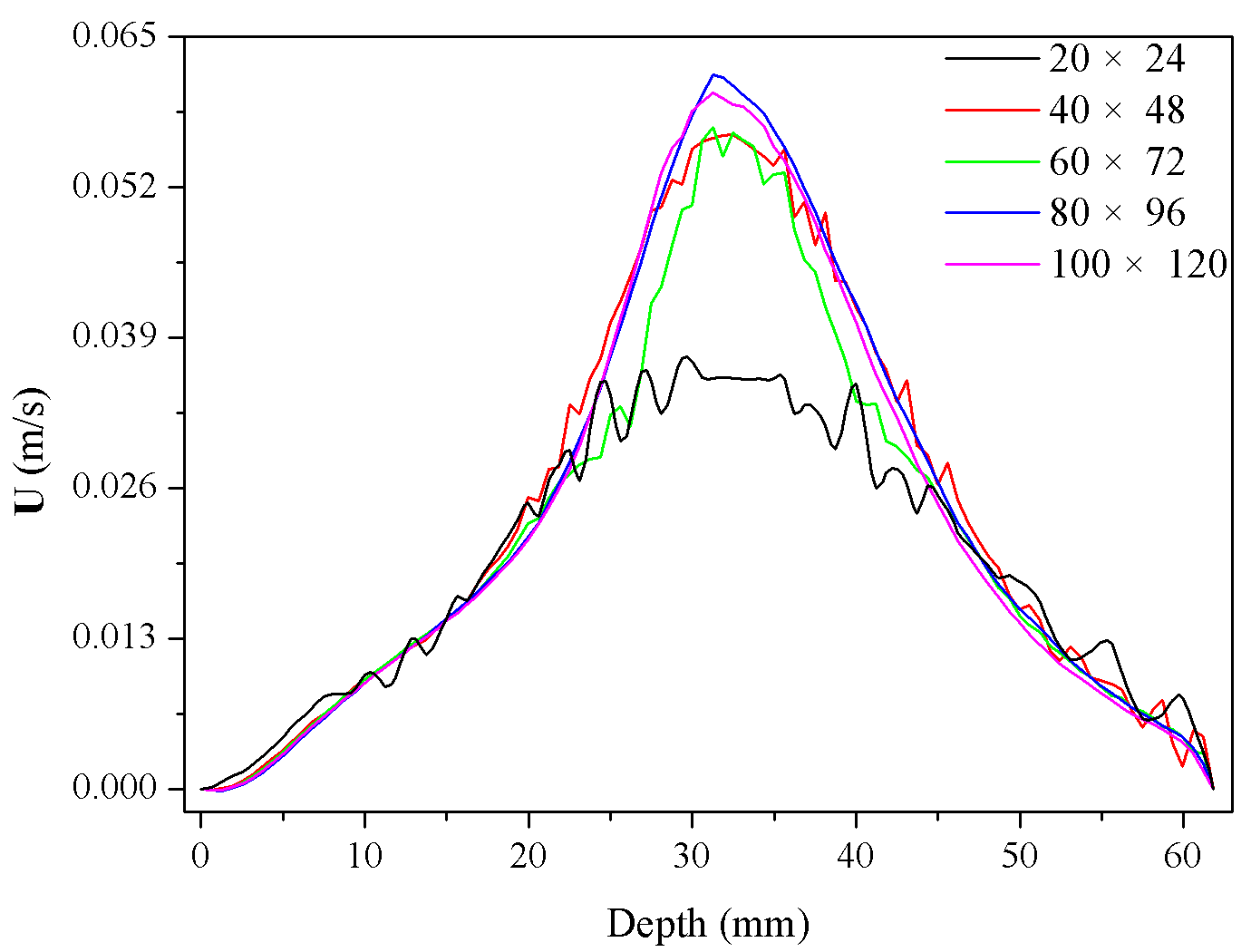}
     		\captionsetup{skip=10pt}
     		\caption{}
     		%\label{fig: parabola surface fit origin 4p contour}
     	\end{subfigure}
     	\hfill
     	\begin{subfigure}[b]{0.485\textwidth}
     		\centering
     		\includegraphics[width=\textwidth]{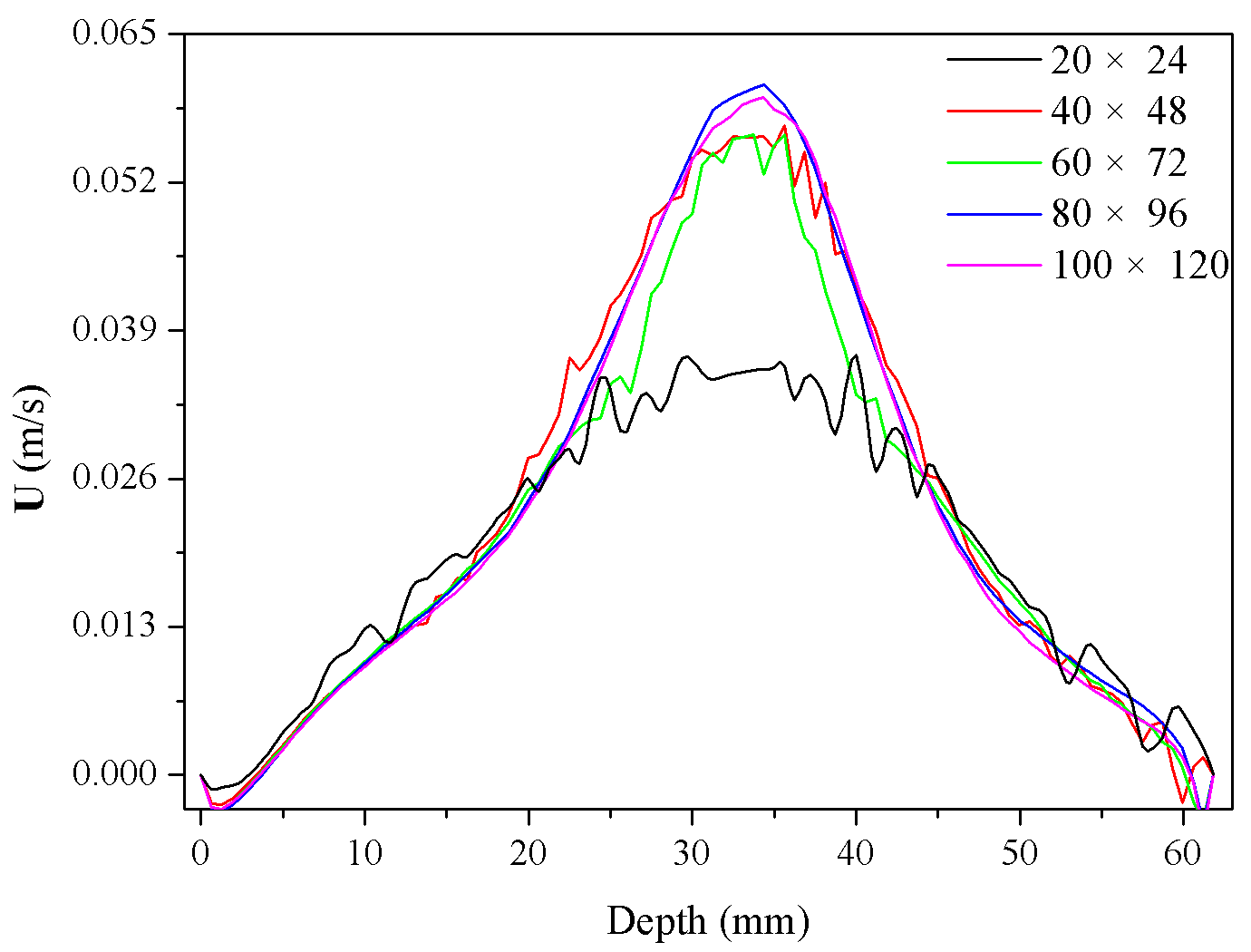}
     		\captionsetup{skip=10pt}
     		\caption{}
     		%\label{fig: parabola surface fit origin 4p curve}
     	\end{subfigure}
     	%\hfill
     	\vspace{10pt}
     	\begin{subfigure}[b]{0.485\textwidth}
     		\centering
     		\includegraphics[width=\textwidth]{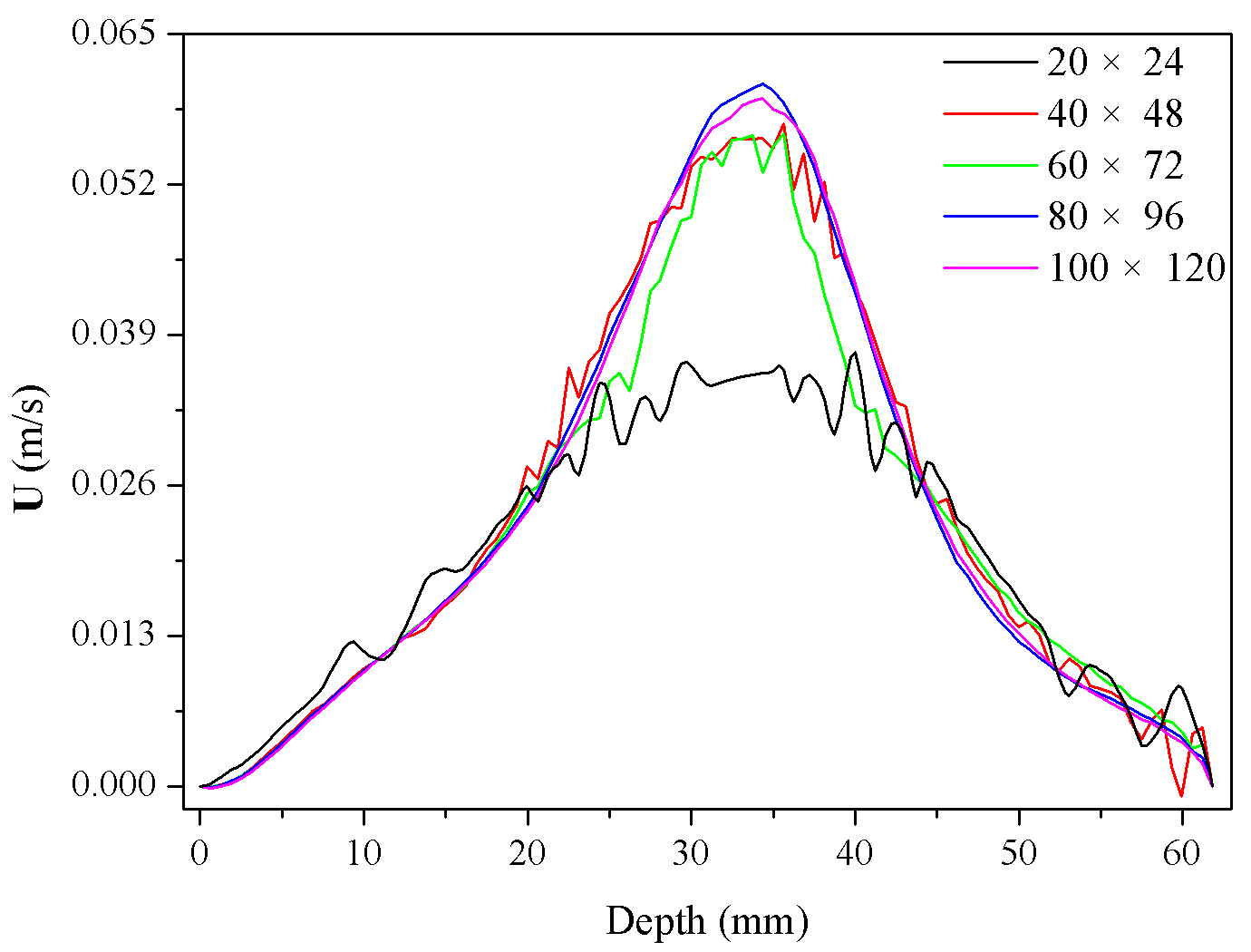}
     		\captionsetup{skip=10pt}
     		\caption{}
     		%\label{fig: parabola surface fit N_4 contour}
     	\end{subfigure}
        \hfill
        \begin{subfigure}[b]{0.485\textwidth}
        	\centering
        	\includegraphics[width=\textwidth]{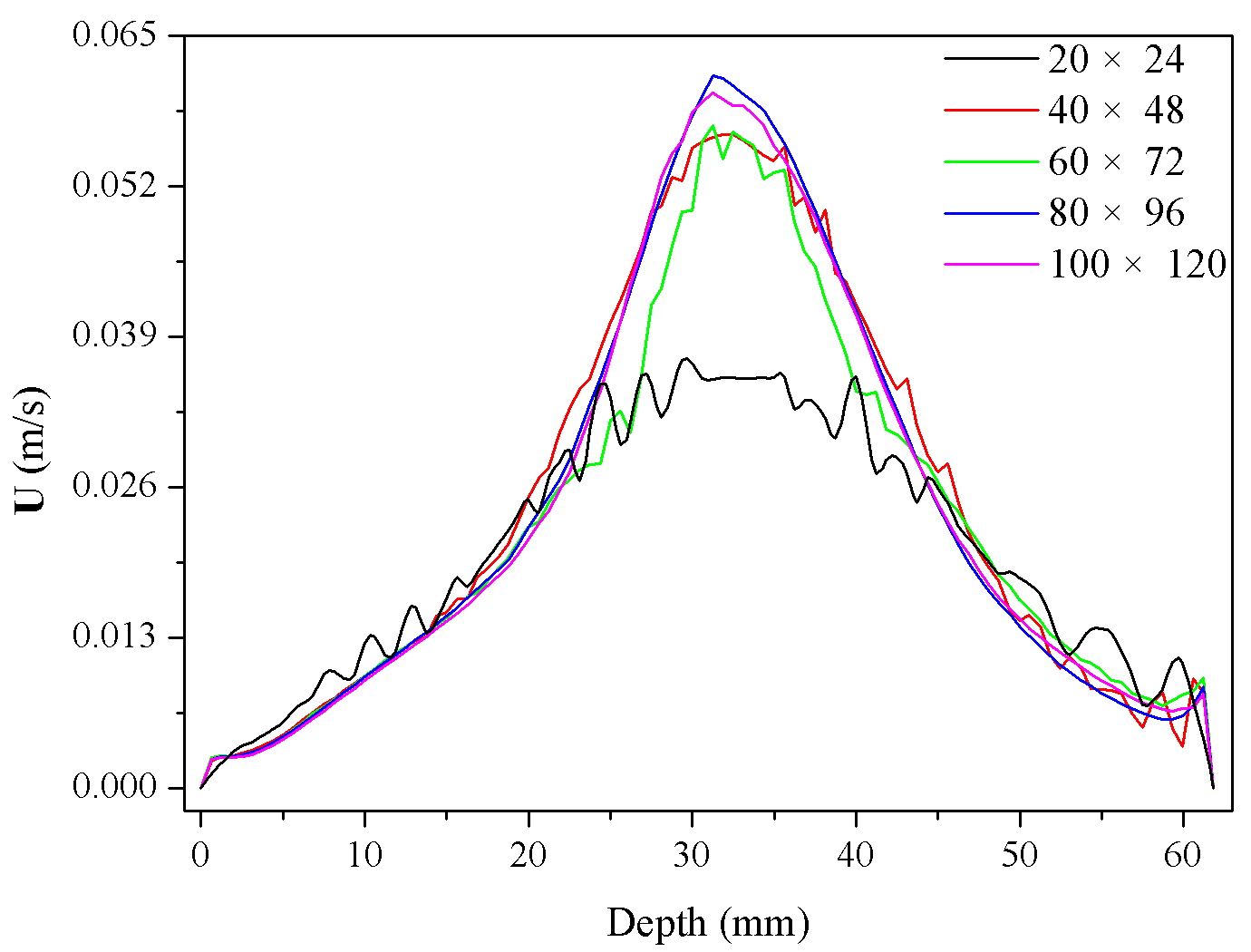}
        	\captionsetup{skip=10pt}
        	\caption{}
     	%\label{fig: parabola surface fit origin 4p curve}
        \end{subfigure}
     	\caption{Distributions of velocities along the beam line of UDV sensor 1, 3, 5 and 7 
	with different grid sizes: (a) sensor 1, (b) sensor 3, (c) sensor 5, (d) sensor 7.}
     	\label{fig:grid_independence_test}
     \end{figure}

     The grid independence test was performed for the case $\textit{\textbf{b}}=0$. Five 
     different grids in the fluid region were used: 
     $20 \times 24 $, $40 \times 48$, $60 \times 72$, $80 \times 96$, $100 \times 120$. 
     Here, the first number (20, 40, 60, 80 and 100) denotes the cell number over the 
     diameter and the second number (24, 48, 72, 96 and 120) means the cell number along the  
     height of the fluid domain. As seen in 
     Fig. \ref{fig:grid_independence_test}, 
     the velocity structure converges for the two highest grid resolutions, so that a
     grid size of $80 \times 96$ is chosen for the further computations.
	
	\begin{figure}[htbp]
		\centering
		\begin{subfigure}[b]{0.39\textwidth}
			\centering
			\includegraphics[width=\textwidth]{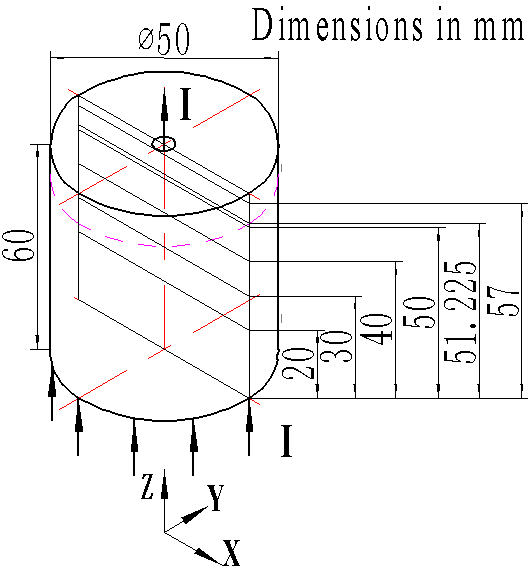}
			\captionsetup{skip=10pt}
			\caption{}
			%\label{fig: parabola surface fit origin 4p contour}
		\end{subfigure}
    	\hfill
		\begin{subfigure}[b]{0.6\textwidth}
		\centering
		\includegraphics[width=\textwidth]{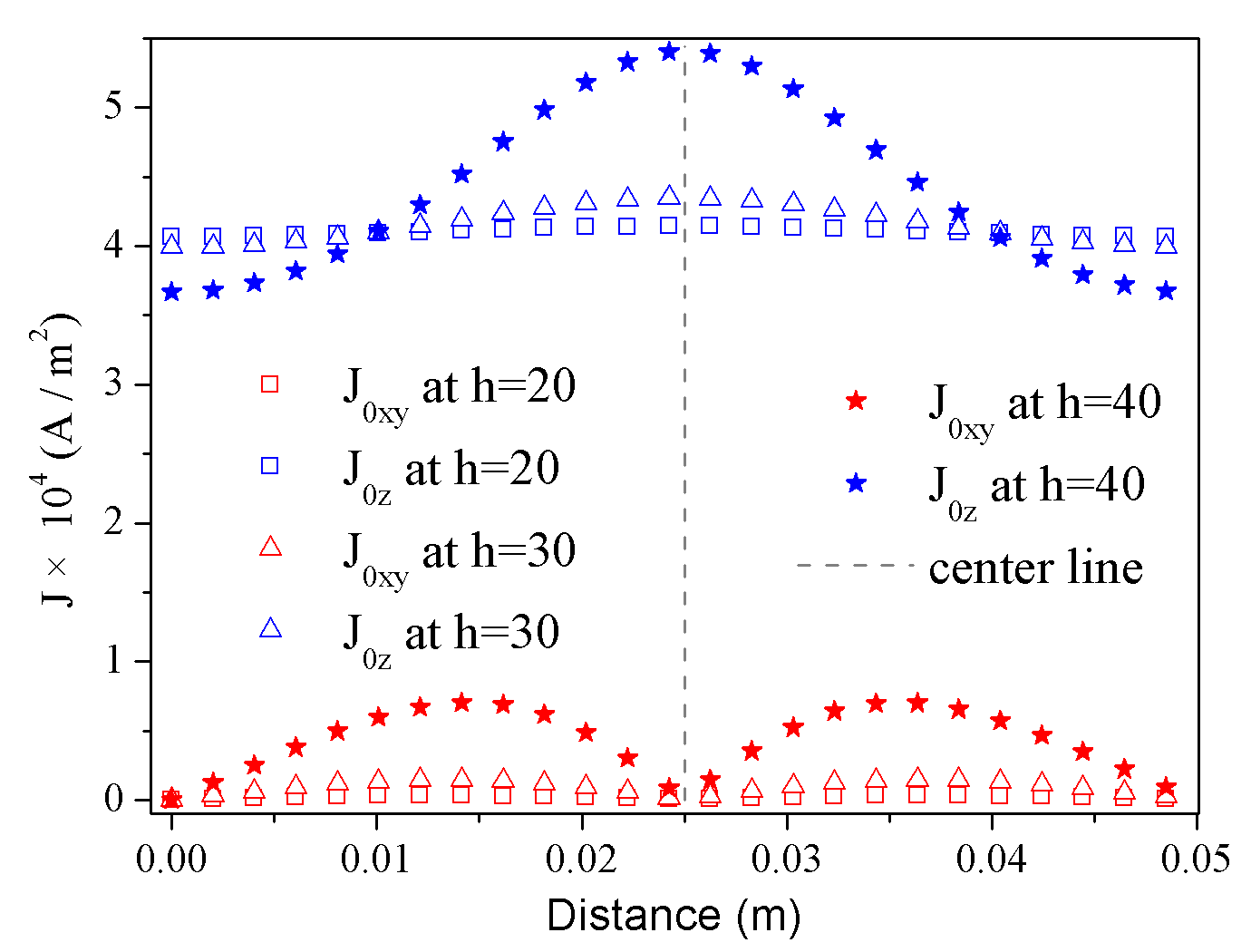}
		\captionsetup{skip=10pt}
		\caption{}
		%\label{fig: parabola surface fit origin 4p curve}
    	\end{subfigure}
    	\vspace{10pt}
        \begin{subfigure}[b]{0.6\textwidth}
    	\centering
    	\includegraphics[width=\textwidth]{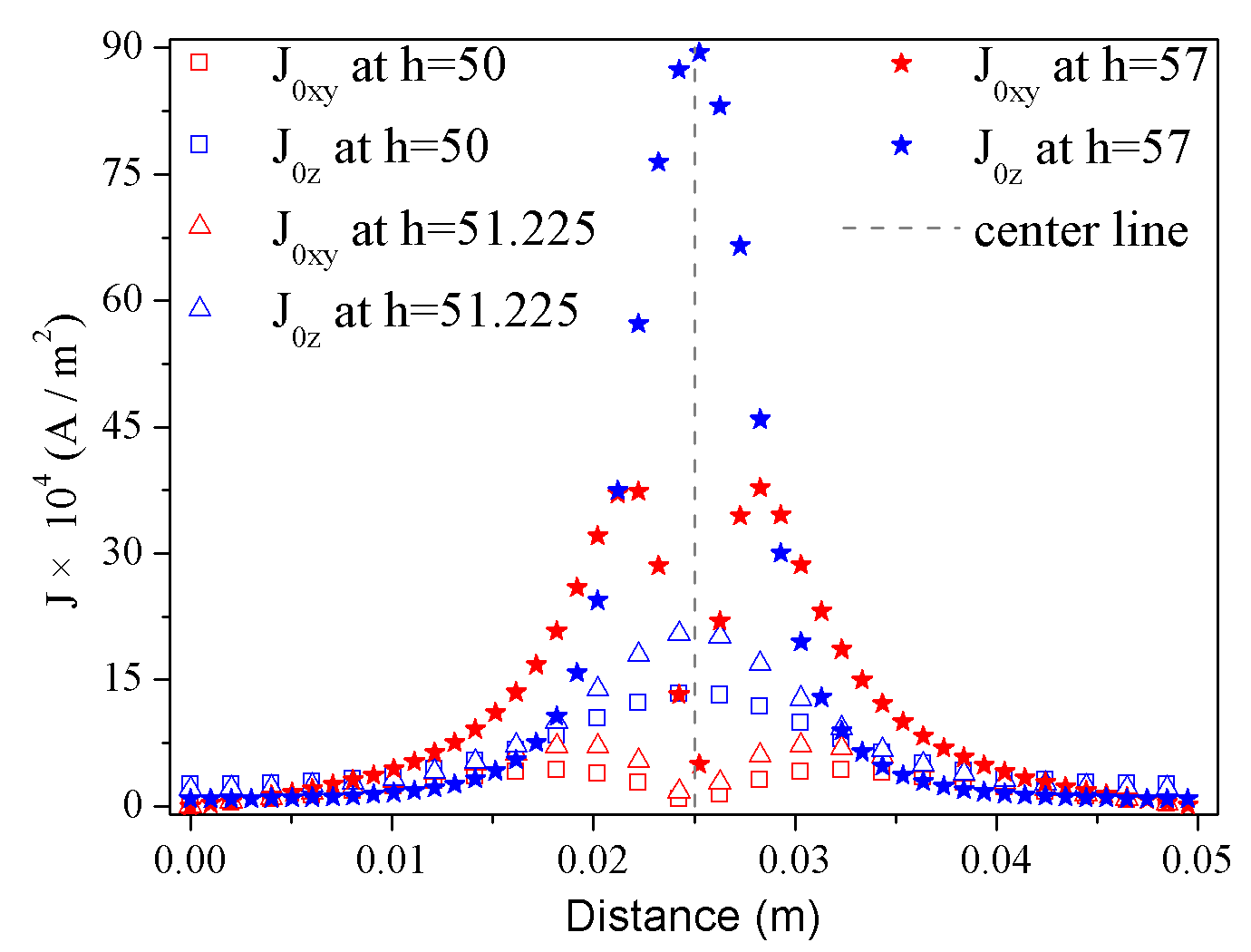}
    	\captionsetup{skip=10pt}
    	\caption{}
    	%\label{fig: parabola surface fit N_4 contour}
        \end{subfigure}
        \caption{Current density distribution at different heights. (a) Positions where the 
	values are sampled. (b) Distribution of current density at $h= 20, 30, 40$\;mm, respectively. 
	(c) Distribution of current density in uppermost part of the container, 
	at $h= 50, 51.225, 57$\;mm, respectively.}
        \label{fig:current_density_comparison}
	\end{figure}

	\begin{figure}[htbp]
		\centering
		\begin{subfigure}[b]{0.375\textwidth}
			\centering
			\includegraphics[width=\textwidth]{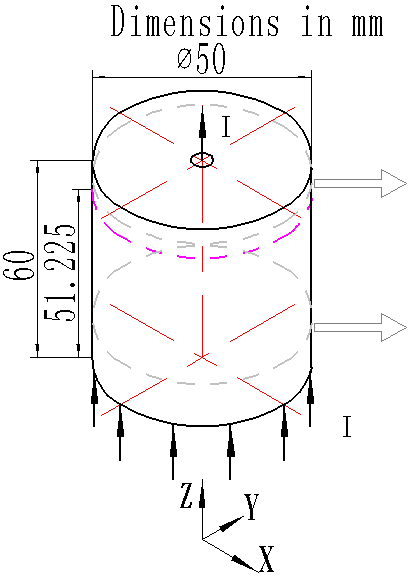}
			\captionsetup{skip=10pt}
			\caption{}
			%\label{fig: parabola surface fit origin 4p contour}
		\end{subfigure}
		\hfill
		\begin{subfigure}[b]{0.195\textwidth}
			\centering
			\includegraphics[width=\textwidth]{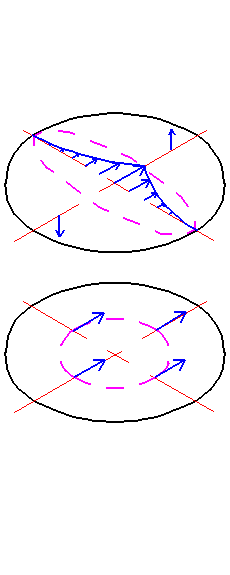}
			\captionsetup{skip=10pt}
			\caption{}
			%\label{fig: parabola surface fit origin 4p curve}
		\end{subfigure}
        \hfill
        \begin{subfigure}[b]{0.195\textwidth}
        	\centering
        	\includegraphics[width=\textwidth]{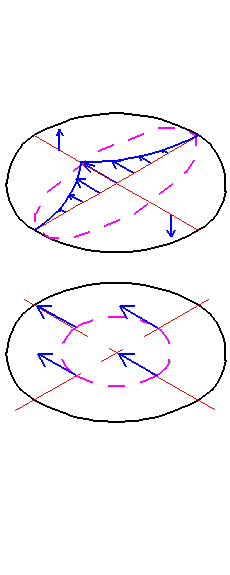}
        	\captionsetup{skip=10pt}
        	\caption{}
        	%\label{fig: parabola surface fit origin 4p curve}
        \end{subfigure}
        \hfill
        \begin{subfigure}[b]{0.195\textwidth}
        	\centering
        	\includegraphics[width=\textwidth]{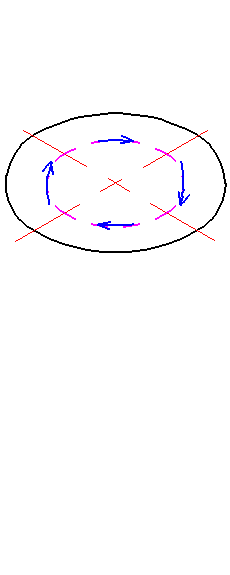}
        	\captionsetup{skip=10pt}
        	\caption{}
        	%\label{fig: parabola surface fit origin 4p curve}
        \end{subfigure}
		\caption{Schematic diagram of the distribution of the Lorentz force on different cross-sections: (a)  Schematic of the fluid domain. (b) The dominant component of the Lorentz force ($J_0 \times b_{x}$), above and below
		 $h=51.225$\;mm. (c) The dominant component of the Lorentz force ($J_0 \times b_{y}$), 
		 above and below $h=51.225$\;mm.
		  (d) The dominant component of the Lorentz force ($J_0 \times b_{z}$).}
		\label{fig:general_distribution_of_the_Lorentz_force}
	\end{figure}

    Figure \ref{fig:current_density_comparison} shows the distribution of the current density along the $x$ 
    axis at different heights. As shown in Fig. \ref{fig:current_density_comparison}\hspace{1.5pt}b and c, 
    the cross-section at $h=51.225$\;mm  (indicated by the 
    pink dashed circle in Fig. \ref{fig:current_density_comparison}\hspace{1.5pt}a) 
    represents a sort of a watershed, which 
    separates the region $z<51.225$\;mm, where the $z$ component of the current density is 
    always larger than the total horizontal component from the region 
    $z>51.225$\;mm, where the $z$ component is not the dominant one.

    Figure \ref{fig:general_distribution_of_the_Lorentz_force} shows some typical distributions of 
    the Lorentz force on different cross-sections, as supposed to be generated by the current 
    and one external magnetic component. Take Fig. \ref{fig:general_distribution_of_the_Lorentz_force}\hspace{1.5pt}b as an    example: the upper sub-plot shows the distribution of the Lorentz force on a cross-section with 
    $z>51.225$\;mm for the case of a pure $b_x$ field. Because in the core 
    area $J_{0z}$ is much stronger than $J_{0xy}$, the Lorentz force 
    points in the positive $y$ direction. However, in the surrounding area, $J_{0xy}$ is much stronger than $J_{0z}$, 
    so the Lorentz force for $y>0$ points up, while it points down for $y<0$. 
    The lower sub-plot illustrates the Lorentz force for $z<51.225$\;mm, i.e. below the watershed,  where on 
    any horizontal cross-section $J_{0z}$ is the strongest component, 
    so that the Lorentz force always points in positive $y$ direction. Similar effects are visible in the 
    two sub-plots 
    of Fig.  \ref{fig:general_distribution_of_the_Lorentz_force}\hspace{1.5pt}c, which illustrate 
    the Lorentz force generated by $\textit{J}_0$ and a pure $b_y$ component. Assuming a pure $b_z$,
    Fig. \ref{fig:general_distribution_of_the_Lorentz_force}\hspace{1.5pt}d shows that 
    only a azimuthal Lorentz force can be generated. It is well known from former experiments \cite{Woods1971,Bojarevics1983} and theory \cite{Millere1980,Davidson1999,Davidson2000a} that such a force drives easily an oscillating \cite{Mandrykin2019} or swirling flow.
Considering all the three sub-plots (Fig. \ref{fig:general_distribution_of_the_Lorentz_force}\hspace{1.5pt}b,c,d together, we see that 
    the Lorentz force parts due to $b_x$ and $b_y$ can partly cancel the contribution due to $b_z$.
    \begin{figure}[htbp]
    	\centering
    	\begin{subfigure}[b]{0.485\textwidth}
    		\centering
    		\includegraphics[width=\textwidth]{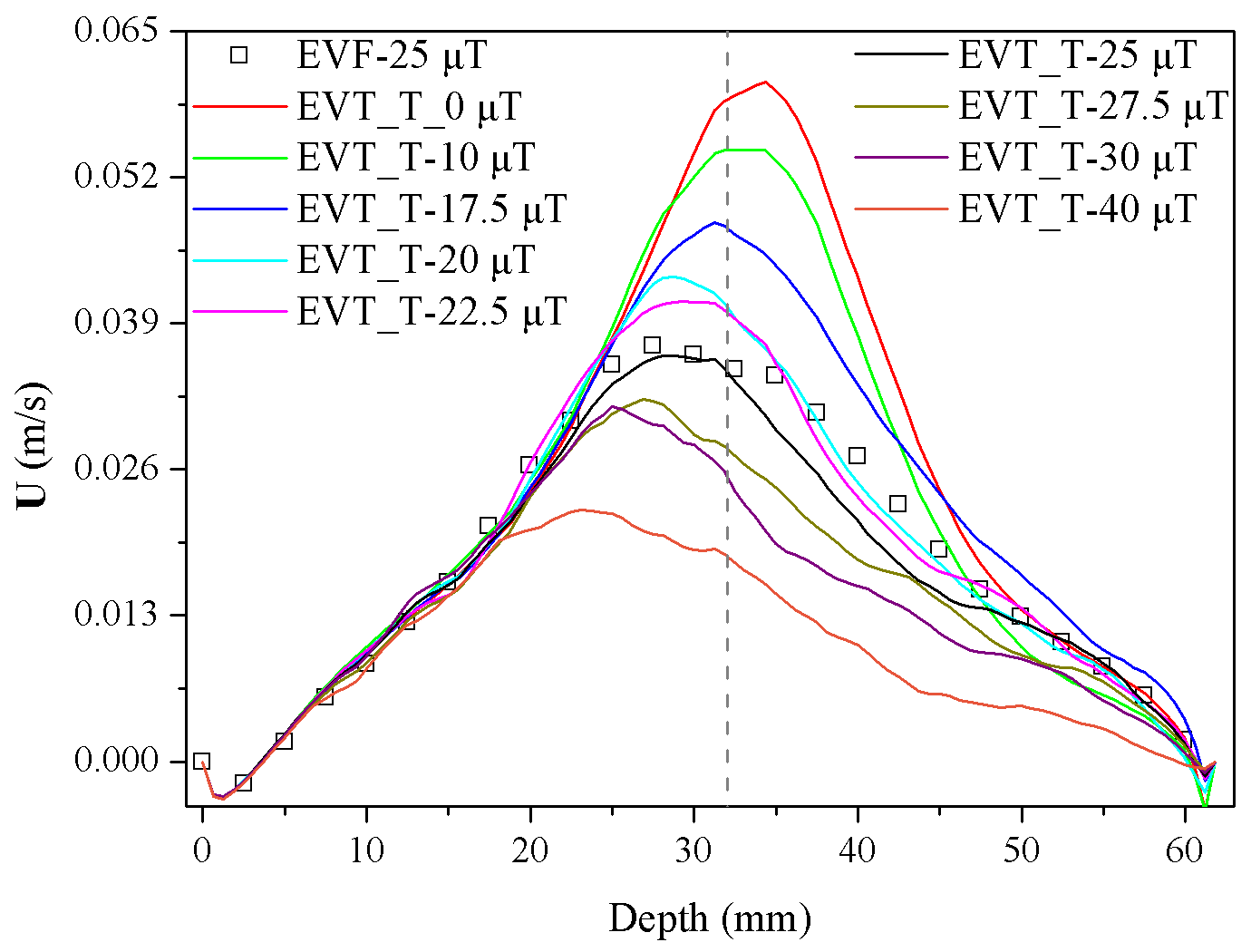}
    		\captionsetup{skip=10pt}
    		\caption{}
    		%\label{fig: parabola surface fit origin 4p contour}
    	\end{subfigure}
    	\hfill
    	\begin{subfigure}[b]{0.485\textwidth}
    		\centering
    		\includegraphics[width=\textwidth]{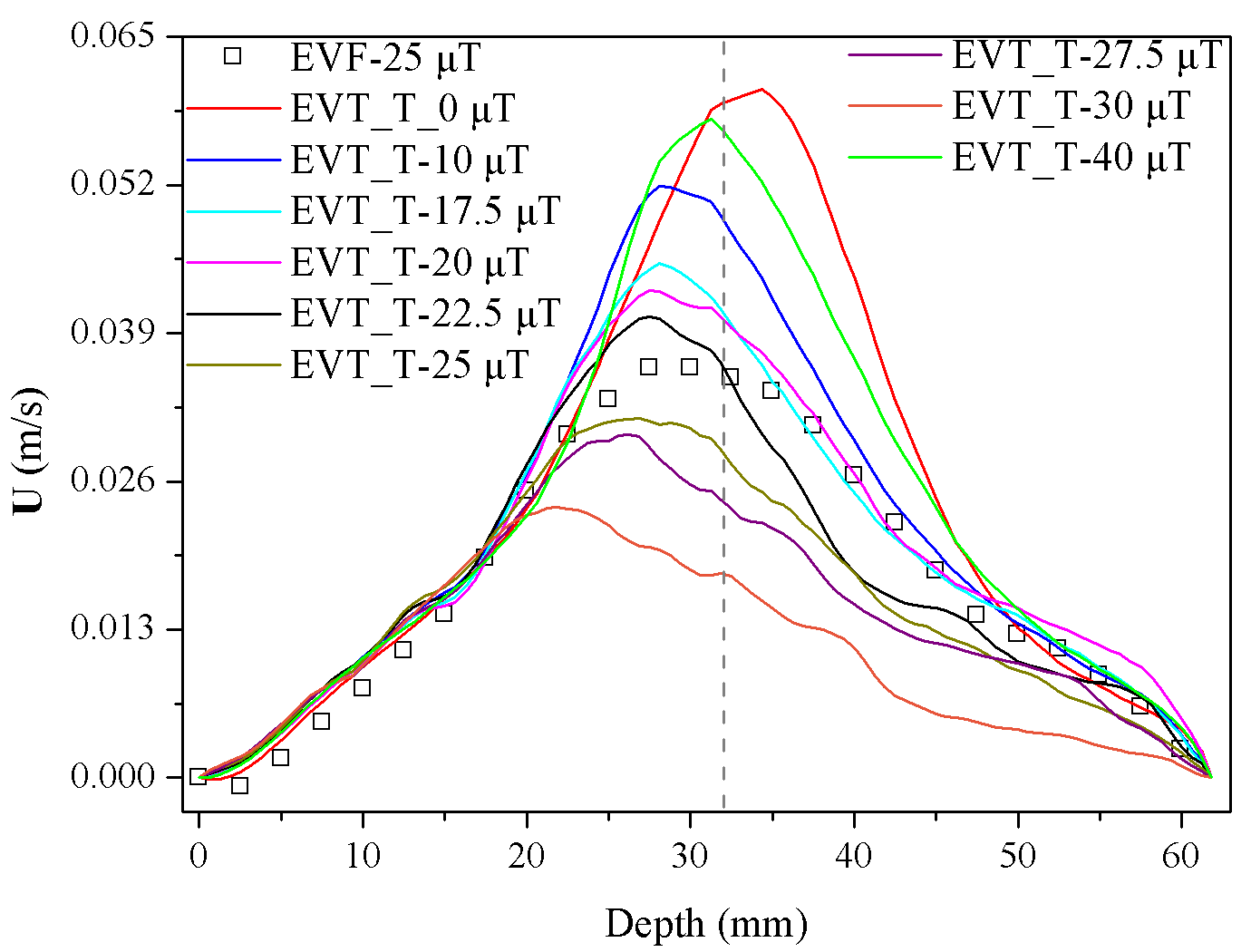}
    		\captionsetup{skip=10pt}
    		\caption{}
    		%\label{fig: parabola surface fit origin 4p curve}
    	\end{subfigure}
    	%\hfill
    	\vspace{10pt}
    	\begin{subfigure}[b]{0.485\textwidth}
    		\centering
    		\includegraphics[width=\textwidth]{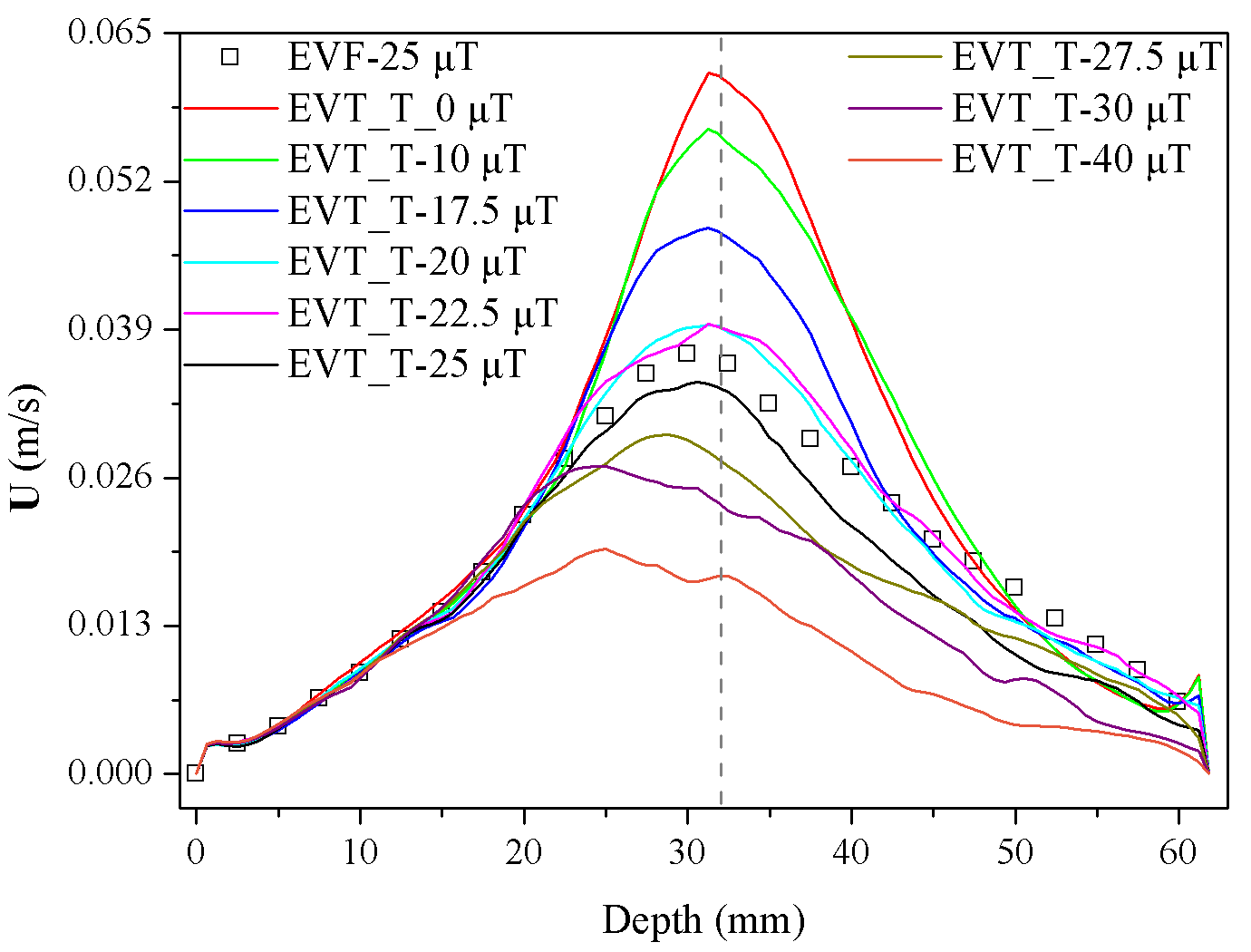}
    		\captionsetup{skip=10pt}
    		\caption{}
    		%\label{fig: parabola surface fit N_4 contour}
    	\end{subfigure}
        \hfill
        \begin{subfigure}[b]{0.485\textwidth}
        	\centering
        	\includegraphics[width=\textwidth]{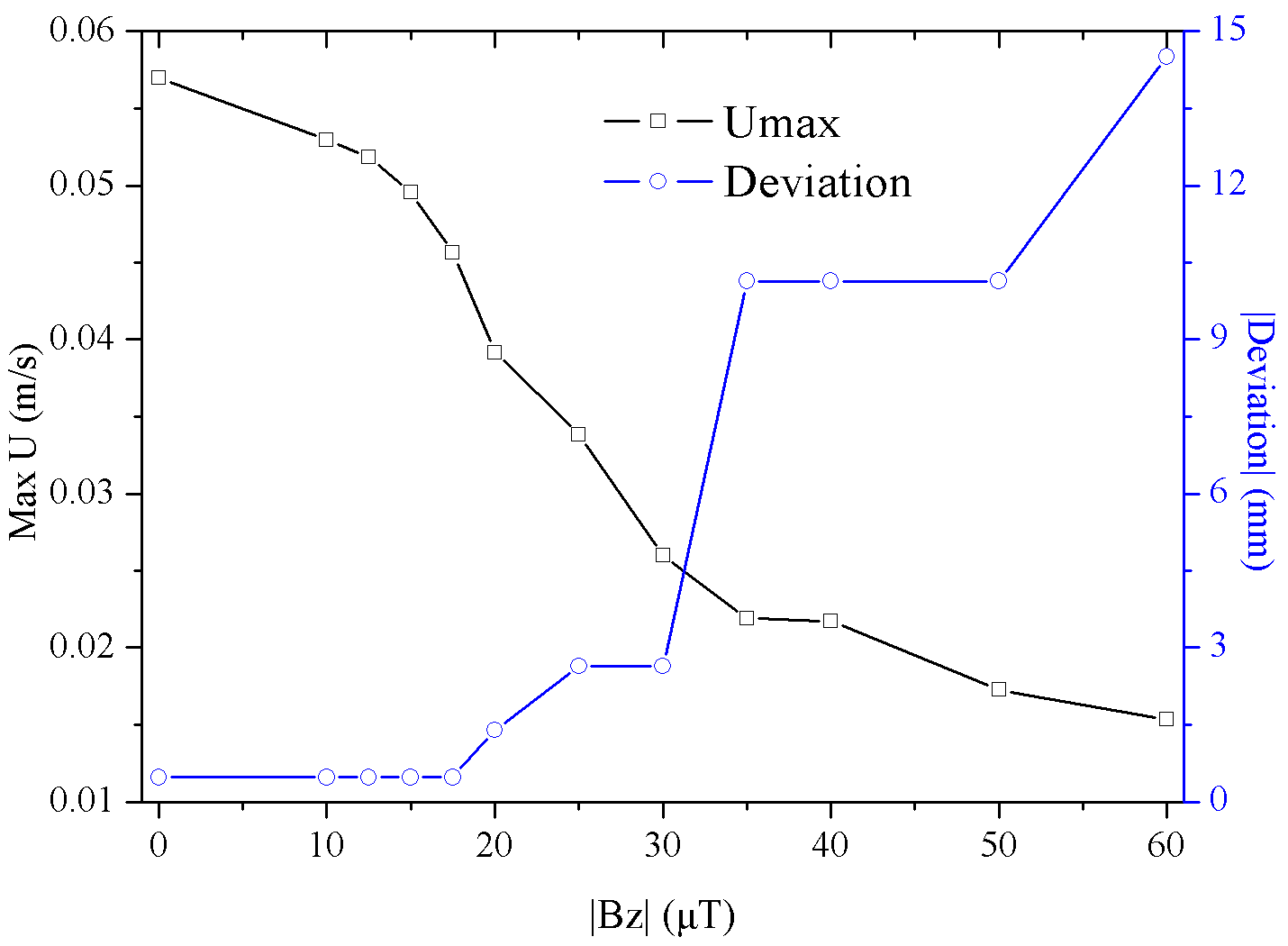}
        	\captionsetup{skip=10pt}
        	\caption{}
        	%\label{fig: parabola surface fit origin 4p curve}
        \end{subfigure}
        \caption{Distributions of velocities along the beam lines of sensor 3, 5 and 7 with different magnitudes of the external magnetic field: (a) sensor 3. (b) sensor 5. (c) sensor 7. (d) Deviation of the velocity peak from the grey dashed line, and maximum velocity for sensor 3.}
        \label{fig:velocity_distribution_with_homogeneous_magnetic_field}
    \end{figure}

    We discuss now the velocity fields that result from the Lorentz forces considered above.
    Figure \ref{fig:velocity_distribution_with_homogeneous_magnetic_field} shows the
    velocity profiles as simulated along the beam line of the UDV sensors 3, 5 and 7. 
    The legends ``EVF"  and 
   ``EVF\_T" indicate numerical results without and with the influence of buoyancy, 
    respectively. Comparing the two settings, for the special case of an applied $b_z=-25.51$\;$\mu$T 
   and $b_x=b_y=0$, Fig. \ref{fig:velocity_distribution_with_homogeneous_magnetic_field}\;b shows that the magnitudes and distributions of the velocity are slightly 
    different. Since the actual value of $b_z$ in the related region close to the lid is not exactly known, we 
    investigate here the effect of applying different values. Evidently, when we 
    gradually vary $b_z$ from zero to $-40$\;$\mu$T, we obtain quite different
    velocity distributions. Typically, with incrasing magnetic field amplitude, the simulated maximum velocity 
    decreases, and its position moves to the left. Figure \ref{fig:velocity_distribution_with_homogeneous_magnetic_field}\;d shows the maximum velocity and its deviation (the absolute value) from the centre line of the cylindrical container (gray dashed line) when $b_z$ varies from $0$ to $-60\;\mu$T in the simulations. We see, in particular, that the maximum velocity 
    is drastically weakened between $b_z=-17.5$\;$\mu$T and $-30$\;$\mu$T. Further below we will
    suggest that the
    measured velocity profiles is well compatible with the presence of 
    an $b_z$ at around $-25$\;$\mu$T in the relevant region close to the 
    upper electrode.
    
    \begin{figure}[H]
    	\centering
    	\begin{subfigure}[b]{0.99\textwidth}
    		\centering
    		\includegraphics[width=\textwidth]{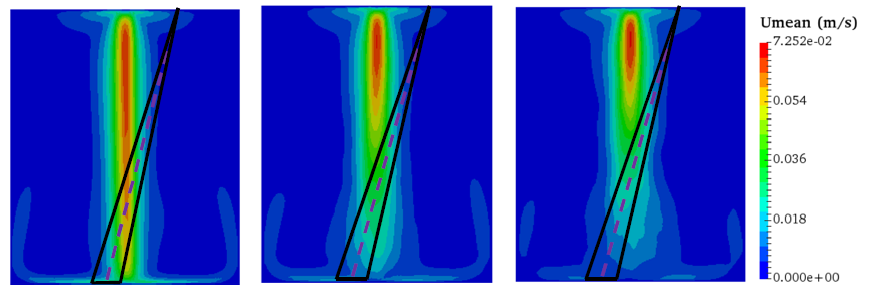}
%    		\captionsetup{skip=10pt}
    		%\caption{four points (Origin\_parabola)}
    		%\label{fig: parabola surface fit origin 4p contour}
    	\end{subfigure}   
    	\vspace{15pt}
    	\begin{subfigure}[b]{0.99\textwidth}
    		\centering
    		\includegraphics[width=\textwidth]{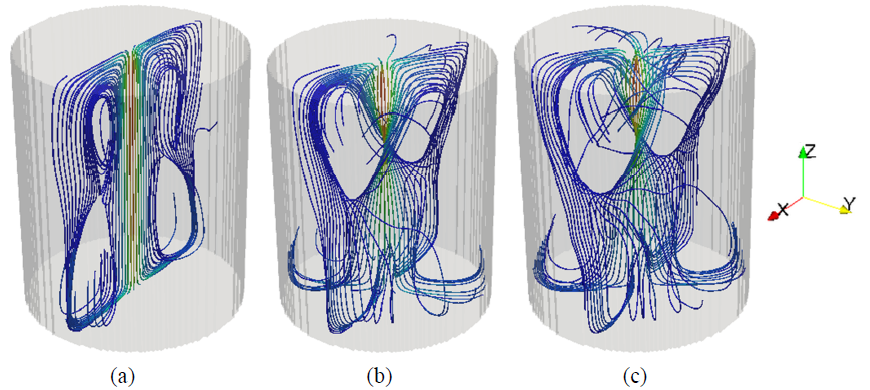}
%    		\captionsetup{skip=10pt}
%    		\caption{}
    		%\label{fig: parabola surface fit N_4 contour}
    	\end{subfigure}	
%    \end{figure}
%    
%    \begin{figure}[H]\ContinuedFloat
%    	\centering
    	\vspace{10pt}
    	\begin{subfigure}[b]{0.99\textwidth}
    		\centering
    		\includegraphics[width=\textwidth]{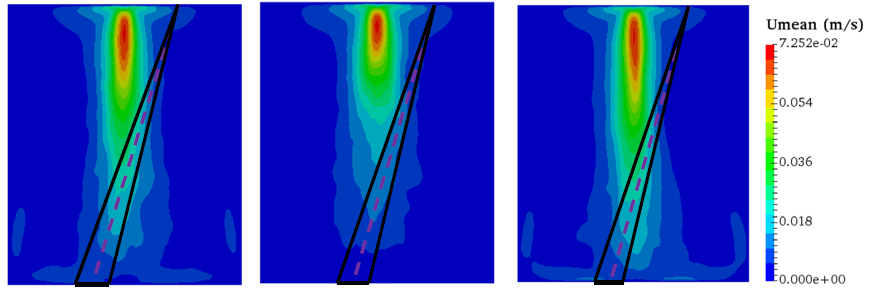}
%    		\captionsetup{skip=10pt}
    		%\caption{four points (Origin\_parabola)}
    		%\label{fig: parabola surface fit origin 4p contour}
    	\end{subfigure} 
    	\vspace{3pt}
    	\begin{subfigure}[b]{0.99\textwidth}
    		\centering
    		\includegraphics[width=\textwidth]{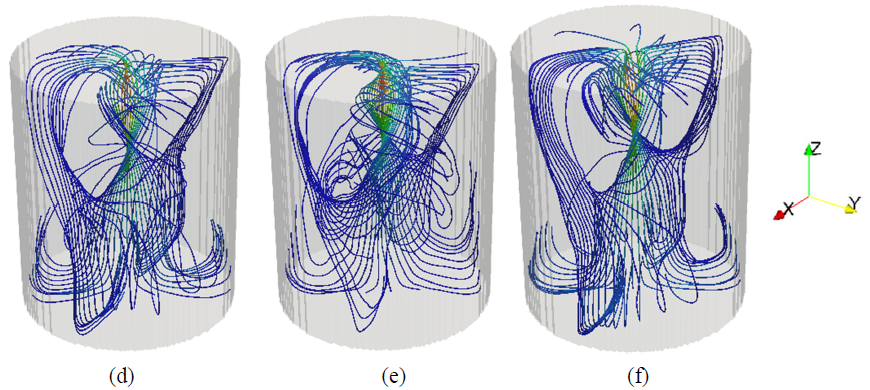}
%    		\captionsetup{skip=10pt}
%    		\caption{}
    		%\label{fig: parabola surface fit N_4 contour}
    	\end{subfigure}
    	\caption{Velocity contours in the ZOX coordinate plane (upper sub-panels), 
	and the corresponding streamlines (lower sub-panels), for 
	 different external magnetic field configurations: (a) $b_z=0$ (b) 
	 $b_z=-20 \hspace{1.5pt} \mu$T (c) $b_z=-25.51 \hspace{1.5pt} \mu$T (d) 
	 $b_z=-30 \hspace{1.5pt} \mu$T (e) $b_z=-40  \hspace{1.5pt} \mu$T (f) 
	 $b_x=15 \hspace{1.5pt} \mu$T, $b_y=33.9 \hspace{1.5pt} \mu$T, $b_z=-25.51 \hspace{1.5pt} \mu$T}
    	\label{fig:velocity_contour_and_streamlines}  
    \end{figure}

Figure \ref{fig:velocity_contour_and_streamlines} illustrates and explains 
this behaviour in more detail. For different magnetic field configurations, 
this figure shows the time-averaged velocity contour in the ZOX coordinate plane
(upper sub-panels) and the corresponding streamlines  (lower sub-panels).
While for $b_z=0$ (Fig. \ref{fig:velocity_contour_and_streamlines}\;a) the jet
occupies the whole centre region down to the bottom, 
its length shrinks drastically when increasing the amplitude of $b_z$.
When  $b_z$ is enhanced from $-20$\;$\mu$T 
to $-40$\;$\mu$T (Fig. \ref{fig:velocity_contour_and_streamlines}\;b-e), 
the length of the jet region decreases 
correspondingly from about 33 percent to about 20 percent of the total height.
When adding (in Fig. \ref{fig:velocity_contour_and_streamlines}\;f)
some horizontal field with $b_x=15$\;$\mu$T and $b_y=33.9$\;$\mu$
to the $b_z=-25.51$\;$\mu$T according to 
Fig. \ref{fig:velocity_contour_and_streamlines}\;c, 
the velocity field remains almost unchanged. Hence, for this kind of EVF,  $b_x$ and $b_y$ 
seem to have only a  minor effect on the magnitude and the distribution of the velocity 
field. 

Note that the black triangles in Fig. \ref{fig:velocity_contour_and_streamlines} with 
purple dashed centre line indicate a conical domain where the experimental 
data are measured by the UDV sensor. It is obvious that this conical domain does 
not pass through the jet domain when the amplitude of the external magnetic field becomes 
stronger than $20$\;$\mu$T. This explains the shifting of the 
''measured'' velocity towards lower depths, as seen in Fig. \ref{fig:velocity_distribution_with_homogeneous_magnetic_field}. Furthermore, the streamlines for $b_z=0$ are essentially 2D; when 
$b_z$ is between $-20$\;$\mu$T and $-30$\;$\mu$T, the two 
vortices at the bottom are still concentrated in the XOZ plane, 
while the top two vortices are already 3D-deformed gradually.
For $b_z=-40$\;$\mu$T, all four vortexes become strongly distorted.

	\begin{figure}[H]
		\centering
		\begin{subfigure}[b]{0.485\textwidth}
			\centering
			\includegraphics[width=\textwidth]{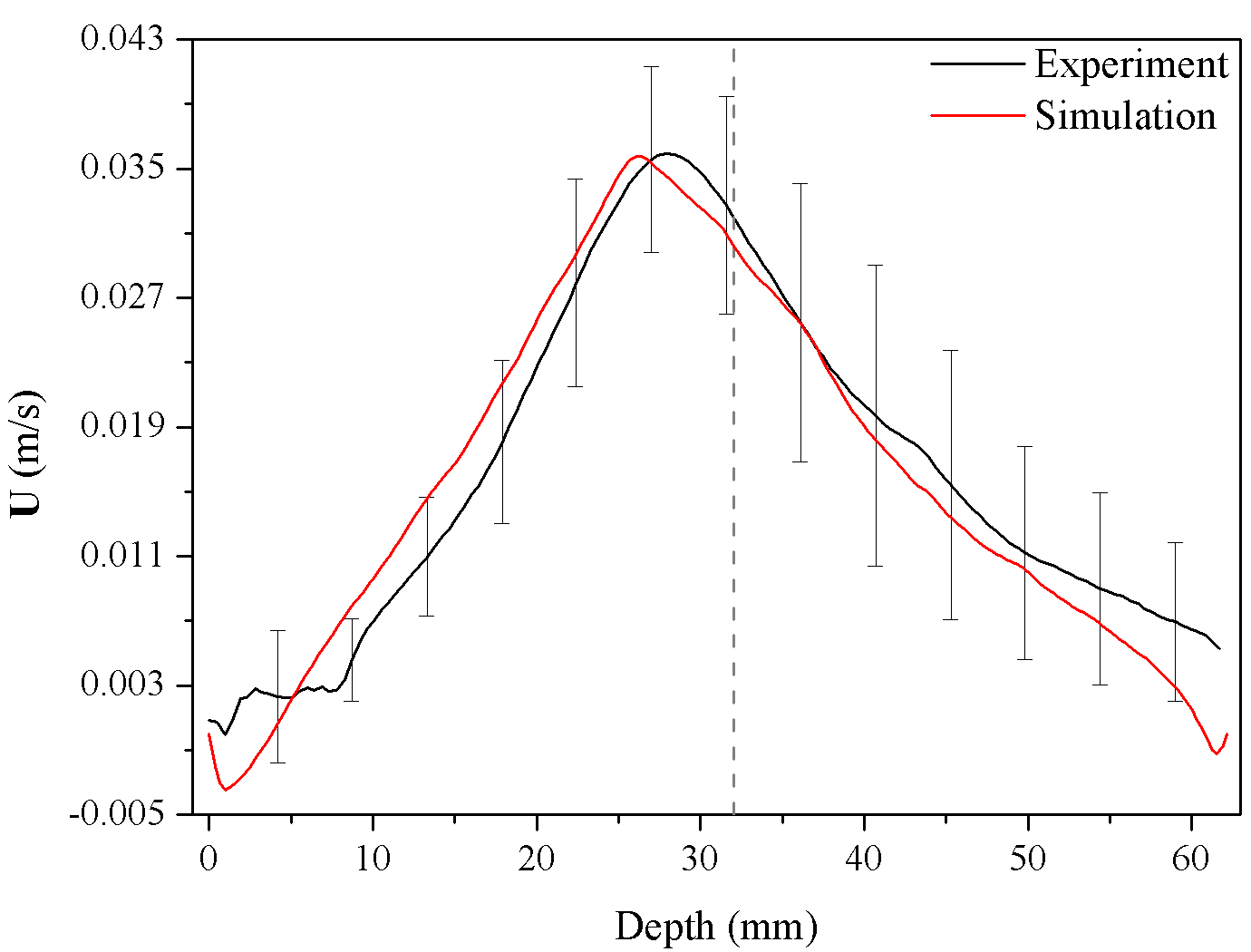}
			\captionsetup{skip=10pt}
			\caption{}
			%\label{fig: parabola surface fit origin 4p contour}
		\end{subfigure}
		\hfill
		\begin{subfigure}[b]{0.485\textwidth}
			\centering
			\includegraphics[width=\textwidth]{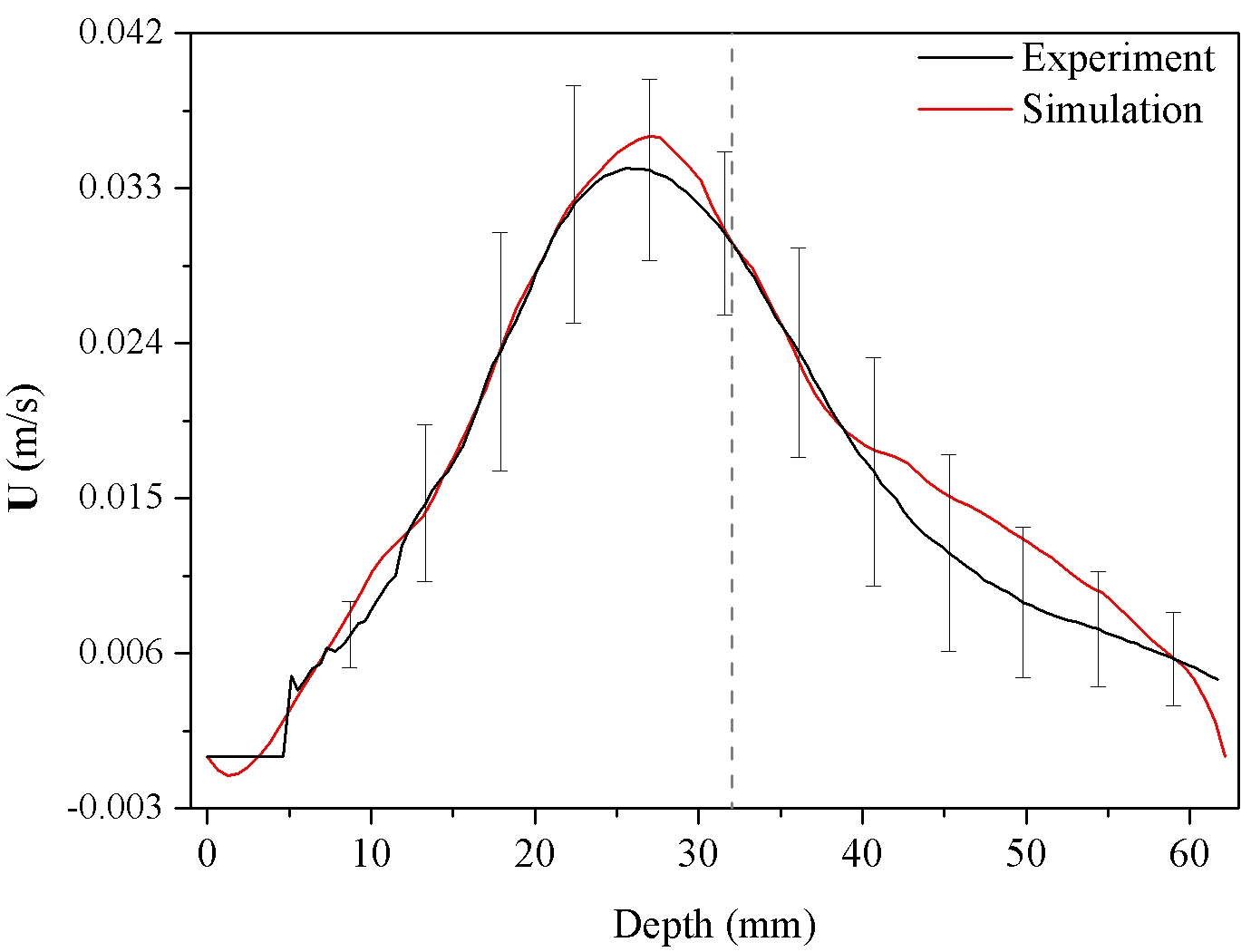}
			\captionsetup{skip=10pt}
			\caption{}
			%\label{fig: parabola surface fit origin 4p curve}
		\end{subfigure}
		%\hfill
		\vspace{10pt}
		\begin{subfigure}[b]{0.485\textwidth}
			\centering
			\includegraphics[width=\textwidth]{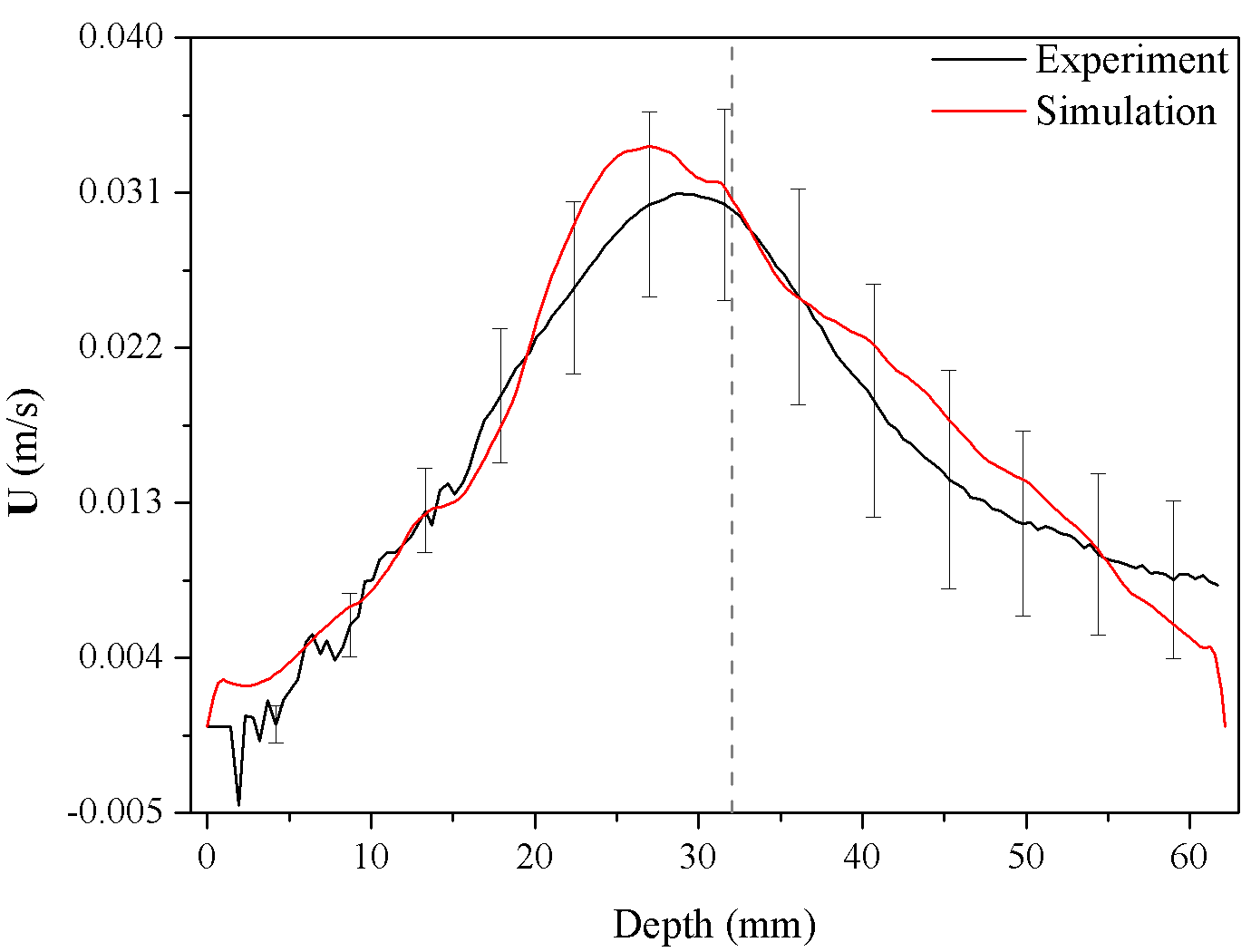}
			\captionsetup{skip=10pt}
			\caption{}
			%\label{fig: parabola surface fit N_4 contour}
		\end{subfigure}
		\caption{Comparison between numerical and experimental velocity 
		profiles along the beams of the UDV sensors for $b_x=15\mu$T, $b_y=33.9\mu$T and $b_z=-25.51\mu$T: (a) sensor 3, 
		(b) sensor 5, (c) sensor 7.}
		\label{fig:velocity_comparison_between_num_and_exp}
	\end{figure}
	
	In Fig. \ref{fig:velocity_comparison_between_num_and_exp}, we show the 
	velocity profiles (with error bars) as measured by UDV along 
	the beams of three sensors (actually, the conical areas as
	shown in Fig. \ref{fig:velocity_contour_and_streamlines}). 
	The red profiles stand for numerical results sampled along the centre line 
	of those UDV sensors, for the special case $b_x=15$\;$\mu$T and $b_y=33.9$\;$\mu$T, $b_z=-25.51$\;$\mu$T (corresponding to Fig. \ref{fig:velocity_contour_and_streamlines}\;f).
	The comparison shows that for this assumed field configuration 
	the numerical results nicely 
	agree with the experiment data. As in Fig. 
	\ref{fig:velocity_distribution_with_homogeneous_magnetic_field}, the 
	grey dashed line in each sub-graph of Fig. \ref{fig:velocity_comparison_between_num_and_exp} 
	shows the position where the centre line of the cylindrical container crosses 
	the centre line of the UDV beam. The high velocity domain corresponds roughly with the region of the jet.
 Because (with increasing $b_z$) the jet domain is shifted upward, 
	the UDV beam  can only pass through the bottom edge area of the high 
	velocity domain. Again, this is the reason why the measured 
	maximum velocity (the peaks of the profiles in Fig. \ref{fig:velocity_distribution_with_homogeneous_magnetic_field} and Fig. \ref{fig:velocity_comparison_between_num_and_exp} ) is shifted away from the 
	centreline of the cylinder.

% ---------------------------- Conclusions -------------------------------
    \section{Conclusions and Outlook}

    In this paper, an experimental and a numerical model of the EVF in a cylindrical container 
    have been built. 
    First, by analysing the current distribution in the liquid metal, we have obtained 
    numerically the Lorentz force that is responsible for generating the EVF. Second, we have found that  the EVF velocity structure depends 
    sensitively on the magnitude of the axial magnetic field $b_z$. 
    Streamlines and 
    contour graphs of the flows have revealed detailed information about the
    physical process that underlies this behaviour. Third, we have compared the numerical results with the 
    experimentally measured UDV data, and found a good correspondence 
    for an assumed $b_z$ of approximately $-25$\;$\mu$T. 
    This is indeed a rather plausible value,
    given the relativly strong variations of the 
    measured external field from the average value of 
    $-16$\;$\mu$T, and in view of some uncertainties 
    related to a possible effect of the current on $b_z$.
    An improvement of the experimental set-up, with a better defined 
    and adjustable $b_z$, might be helpful in supporting 
    and generalising our results. Future work 
    will  concentrate on the understanding of the time-dependence of 
    the fluctuating jet, which becomes noticeable only for stronger $b_z$. 
%    Third, we have compared the numerical 
%    results with the experimentally measured UDV data, and found a good correspondence for an
%    assumed $b_z$ of approximately $-25$\;$\mu$T. Future work will concentrate on the
%    understanding of the time-dependence of the fluctuating jet, which becomes 
%    noticeable only for higher $b_z$. An improvement of the experimental set-up,
%    with a better defined and adjustable $b_z$, might be helpful in supporting 
%    and generalising our results.

% -------------------------- Acknowledgments -----------------------------
    \section*{Acknowledgements}
    This work was supported by Deutsche Forschungsgemeinschaft (DFG, German Research Foundation) under award number 338560565, and by China Scholarship Council (CSC).
    	
% ----------------------------- -- Bib -----------------------------------
%\bibliographystyle{mhd}

\lastpageno
\end{document}